\documentclass[english,12pt]{article}
\usepackage[T1]{fontenc}
\usepackage[latin9]{inputenc}
\usepackage{geometry}
\usepackage{color}
\usepackage{babel}
\usepackage{amsmath}
\usepackage{amsthm}
\usepackage{amssymb}
\usepackage{mleftright}
\usepackage[pdfusetitle,
 bookmarks=true,bookmarksnumbered=false,bookmarksopen=false,
 breaklinks=false,pdfborder={0 0 0},pdfborderstyle={},backref=false,colorlinks=true]
 {hyperref}
\hypersetup{
 linkcolor=vblue, citecolor=vblue}

\makeatletter
\numberwithin{equation}{section}
\theoremstyle{plain}
\newtheorem*{prop*}{\protect\propositionname}
\theoremstyle{plain}
\newtheorem{thm}{\protect\theoremname}
\theoremstyle{plain}
\newtheorem{conjecture}[thm]{\protect\conjecturename}
\theoremstyle{plain}
\newtheorem{prop}[thm]{\protect\propositionname}

\definecolor{vblue}{RGB}{0,0,255}

\allowdisplaybreaks
\makeatother

\providecommand{\conjecturename}{Conjecture}
\providecommand{\propositionname}{Proposition}
\providecommand{\theoremname}{Theorem}

\begin{document}
\title{A Correlational Bound for Eigenvalues of Fermionic 2-Body Operators}
\author{Martin Ravn Christiansen\\
{\footnotesize ISTA (Institute of Science and Technology Austria),
3400 Klosterneuburg, Austria}}
\maketitle
\begin{abstract}
We prove that the eigenvalues of a 2-body operator $\gamma_{2}^{\Psi}$
associated to a fermionic $N$-particle state $\Psi$ are highly constrained
by the structure of the corresponding eigenvectors: If $\Phi=\sum_{k=1}^{\infty}\lambda_{k}u_{k}\wedge v_{k}$
is the canonical form of an eigenvector $\Phi$ with eigenvalue $\Lambda$,
then $\Lambda\leq\mleft(1+\frac{N-2}{2}\sum_{k=1}^{\infty}\lambda_{k}^{4}\mright)^{-1}N$.

We also prove a lower bound on $\sup_{\left\Vert \Psi\right\Vert =1}\left\langle \Phi,\gamma_{2}^{\Psi}\Phi\right\rangle $
for fixed $\Phi$, and state a conjecture motivated by these results.
\end{abstract}

\section{Introduction}

Let $\mathfrak{h}$ be a separable Hilbert space and consider the
space of fermionic $N$-particle states $\bigwedge^{N}\mathfrak{h}$.
Given a normalized state $\Psi\in\bigwedge^{N}\mathfrak{h}$ we define
the $2$-body operator associated to $\Psi$, $\gamma_{2}^{\Psi}:\mathfrak{h}\otimes\mathfrak{h}\rightarrow\mathfrak{h}\otimes\mathfrak{h}$,
by
\begin{equation}
\left\langle \mleft(\varphi_{1}\otimes\varphi_{2}\mright),\gamma_{2}^{\Psi}\mleft(\psi_{1}\otimes\psi_{2}\mright)\right\rangle =\left\langle \Psi,c^{\ast}(\psi_{1})c^{\ast}(\psi_{2})c(\varphi_{2})c(\varphi_{1})\Psi\right\rangle 
\end{equation}
for any $\varphi_{1},\varphi_{2},\psi_{1},\psi_{2}\in\mathfrak{h}$,
where $c^{\ast}(\cdot)$ and $c(\cdot)$ denote the fermionic creation
and annihilation operators.

It is well-known that $\gamma_{2}^{\Psi}$ is a non-negative self-adjoint
operator with $\mathrm{tr}(\gamma_{2}^{\Psi})=N(N-1)$. Yang\cite{Yang-62}
proved that $\gamma_{2}^{\Psi}$ also obeys the upper bound $\gamma_{2}^{\Psi}\leq N$,
which is optimal in the infinite-dimensional case.\smallskip{}

Recently, the relation between the eigenvalues of $\gamma_{2}^{\Psi}$
(and more general $n$-body operators) and correlation/entanglement
of the associated states has attracted some attention, in particular
in the form of universal bounds on the entropy\cite{CarLieReu-16,Lemm-24}
and Hilbert-Schmidt norms \cite{Christiansen-24,Visconti-25} of these.\smallskip{}

In this paper we present a generalization of Yang's bound for the
individual eigenvalues: Since $\gamma_{2}^{\Psi}$ is a self-adjoint
trace-class operator, hence is diagonalizable, Yang's bound can be
rephrased in terms of eigenvalues as saying that every eigenvalue
$\Lambda$ of $\gamma_{2}^{\Psi}$ obeys $\Lambda\leq N$. We will
prove that this bound can be significantly strengthened given additional
information on the associated eigenvector.

\subsubsection*{The Canonical Form of an Anti-Symmetric $2$-Tensor}

To state the result we recall that any eigenvector $\Phi$ of $\gamma_{2}^{\Psi}$
is likewise anti-symmetric and that such $2$-tensors can be represented
in the following manner\footnote{In \cite{CarLieReu-16} this is attributed to Yang\cite{Yang-62}
and Youla\cite{Youla-61}, but Youla's paper itself appears to refer
to this result as ``classical''.}:
\begin{prop*}
Let $\Phi\in\mathfrak{h\wedge h}$ be normalized. Then there exists
mutually orthonormal sets $(u_{k})_{k=1}^{\infty},(v_{k})_{k=1}^{\infty}\subset\mathfrak{h}$,
which jointly span $\mathfrak{h}$, and $(\lambda_{k})_{k=1}^{\infty}\subset\left[0,\infty\right)$
with $\sum_{k=1}^{\infty}\lambda_{k}^{2}=1$ such that
\[
\Phi=\sum_{k=1}^{\infty}\lambda_{k}u_{k}\wedge v_{k}.
\]
\end{prop*}
(This is a consequence of the natural bijection between $\mathfrak{h}\otimes\mathfrak{h}$
and the space of conjugate-linear Hilbert-Schmidt operators on $\mathfrak{h}$.)\smallskip{}

Since tensors of the form $u\wedge v:=\frac{1}{\sqrt{2}}(u\otimes v-v\otimes u)$
constitute the most elementary anti-symmetric $2$-tensors, the proposition
guarantees the existence of a basis for $\mathfrak{h}$ which is particularly
``adapted'' to $\Phi$.

In this context the distribution of the coefficients $(\lambda_{k})_{k=1}^{\infty}$
can be seen as a measure of inherent correlation of $\Phi$: If $(\lambda_{k})_{k=1}^{\infty}$
is concentrated at only a few values $k_{1},\ldots,k_{n}\in\mathbb{N}$,
then $\Phi$ can evidently be approximated well by a sum of only $n$
elementary tensors $u\wedge v$, whereas if $(\lambda_{k})_{k=1}^{\infty}$
is highly delocalized this is not possible.\smallskip{}

To make the canonical form more concrete, let us give an explicit
example: Consider $\mathfrak{h}=L^{2}([0,1]^{3};\mathbb{C}^{2})$
with $\mathbb{C}^{2}=\mathrm{span}\{\left|\uparrow\right\rangle ,\left|\downarrow\right\rangle \}$.
Then $\mathfrak{h}\otimes\mathfrak{h}$ can be identified with $L^{2}([0,1]^{3}\times[0,1]^{3};\mathbb{C}^{2}\otimes\mathbb{C}^{2})$,
and a notable subset of $\mathfrak{h}\wedge\mathfrak{h}$ within this
space is given by the functions $\Phi:[0,1]^{3}\times[0,1]^{3}\rightarrow\mathbb{C}^{2}\otimes\mathbb{C}^{2}$
of the form
\begin{equation}
\Phi\mleft(x,y\mright)=\varphi\mleft(x-y\mright)\left|\uparrow\right\rangle \wedge\left|\downarrow\right\rangle \label{eq:ExampleState}
\end{equation}
for even functions $\varphi:[0,1]^{3}\rightarrow\mathbb{C}$. Wavefunctions
of this form are precisely those describing two particles with total
momentum $0$ in the singlet spin state $\left|\uparrow\right\rangle \wedge\left|\downarrow\right\rangle $.

The canonical form of such $\Phi$ can be determined by Fourier expansion:
If $\varphi(x)=\sum_{k\in\mathbb{Z}^{3}}\hat{\varphi}_{k}e^{2\pi ik\cdot x}$
is the Fourier expansion of $\varphi$, then evenness implies $\hat{\varphi}_{k}=\hat{\varphi}_{-k}$.
Consequently
\begin{align}
\Phi(x,y) & =\frac{1}{\sqrt{2}}\sum_{k\in\mathbb{Z}^{3}}\hat{\varphi}_{k}\mleft(e^{2\pi ik\cdot x}e^{-2\pi ik\cdot y}\left|\uparrow\right\rangle \otimes\left|\downarrow\right\rangle -e^{2\pi ik\cdot x}e^{-2\pi ik\cdot y}\left|\downarrow\right\rangle \otimes\left|\uparrow\right\rangle \mright)\nonumber \\
 & =\frac{1}{\sqrt{2}}\sum_{k\in\mathbb{Z}^{3}}\hat{\varphi}_{k}\mleft(e^{2\pi ik\cdot x}e^{-2\pi ik\cdot y}\left|\uparrow\right\rangle \otimes\left|\downarrow\right\rangle -e^{-2\pi ik\cdot x}e^{2\pi ik\cdot y}\left|\downarrow\right\rangle \otimes\left|\uparrow\right\rangle \mright)\\
 & =\sum_{k\in\mathbb{Z}^{3}}\hat{\varphi}_{k}\mleft(e^{2\pi ik\cdot(\cdot)}\left|\uparrow\right\rangle \wedge e^{-2\pi ik\cdot(\cdot)}\left|\downarrow\right\rangle \mright)(x,y)\nonumber 
\end{align}
from which the canonical form of $\Phi$ can be read off to be given
by $u_{k}(x)=e^{2\pi ik\cdot x}\left|\uparrow\right\rangle $, $v_{k}(x)=e^{-2\pi ik\cdot x}\left|\downarrow\right\rangle $
and $\lambda_{k}=\hat{\varphi}_{k}$ (indexed over $\mathbb{Z}^{3}$
instead of $\mathbb{N}$).

\subsection{Main Results}

Our main result is the following:
\begin{thm}
\label{them:CorrelationalBound}Let $\Psi\in\bigwedge^{N}\mathfrak{h}$
be normalized, $N\geq2$, and let $\Phi\in\mathfrak{h}\wedge\mathfrak{h}$
be a normalized eigenvector of $\gamma_{2}^{\Psi}$ with eigenvalue
$\Lambda$ and canonical form $\Phi=\sum_{k=1}^{\infty}\lambda_{k}u_{k}\wedge v_{k}$.
Then
\[
\Lambda\leq\frac{N}{1+\frac{N-2}{2}\sum_{k=1}^{\infty}\lambda_{k}^{4}}.
\]
\end{thm}

As $\sum_{k=1}^{\infty}\lambda_{k}^{2}=1$, the sum $\sum_{k=1}^{\infty}\lambda_{k}^{4}$
which appears on the right-hand side is a measure of delocalization
for $(\lambda_{k})_{k=1}^{\infty}$, which as mentioned above can
be considered a measure of correlation of $\Phi$. It is for this
reason that we refer to this as a \textit{correlational }bound, and
the inequality can be seen as saying that ``large eigenvalues $\Lambda$
necessitates highly correlated eigenvectors $\Phi$''.\smallskip{}

The inequality clearly implies Yang's bound $\Lambda\leq N$. What
is rather remarkable is that it is also good ``far away'' from the
optimum of this - at the most extreme end, if $\Phi=u\wedge v$ is
an eigenvector of some 2-body operator then the inequality yields
$\Lambda\leq\frac{N}{1+\frac{N-2}{2}}=2$, which is in fact attained
for Slater states.

\subsubsection*{An Almost Converse Statement for Highly Correlated States}

For a given $\Phi=\sum_{k=1}^{\infty}\lambda_{k}u_{k}\wedge v_{k}$
we let $\lambda_{\max}=\max_{k\in\mathbb{N}}\lambda_{k}$. As $\sum_{k=1}^{\infty}\lambda_{k}^{4}\leq\lambda_{\max}^{2}\sum_{k=1}^{\infty}\lambda_{k}^{2}=\lambda_{\max}^{2}$,
expanding the right-hand side of the inequality of Theorem \ref{them:CorrelationalBound}
with respect to $\sum_{k=1}^{\infty}\lambda_{k}^{4}$ in particular
yields
\begin{equation}
\Lambda\leq N\mleft(1-\frac{N-2}{2}\sum_{k=1}^{\infty}\lambda_{k}^{4}+O(N^{2}\lambda_{\max}^{4})\mright).\label{eq:AsymptoticUpperBound}
\end{equation}
This can be viewed as an asymptotic statement regarding the highly
correlated regime $N\lambda_{\max}^{2}\ll1$, and it is natural to
ask if this is optimal in some sense. To that end we will also prove
the following:
\begin{thm}
\label{them:LowerBound}Let $\Phi\in\mathfrak{h}\wedge\mathfrak{h}$
be normalized with canonical form $\Phi=\sum_{k=1}^{\infty}\lambda_{k}u_{k}\wedge v_{k}$.
Then for every $N\in2\mathbb{N}$ with $N\lambda_{\max}^{2}\leq1$
there exists a normalized state $\Psi\in\bigwedge^{N}\mathfrak{h}$
such that
\[
\left\langle \Phi,\gamma_{2}^{\Psi}\Phi\right\rangle \geq N\mleft(1-\frac{N-2}{2}\sum_{k=1}^{\infty}\lambda_{k}^{4}-\frac{1}{2}(N\lambda_{\max}^{2})^{2}\mright).
\]
\end{thm}

This is not quite a true converse statement, as we do not assert that
$\Psi$ can be chosen such that $\Phi$ is also an eigenvector of
$\gamma_{2}^{\Psi}$, which is in turn needed for our proof of Theorem
\ref{them:CorrelationalBound}, hence for equation (\ref{eq:AsymptoticUpperBound}).\smallskip{}

We remark however that the trial states we will construct for Theorem
\ref{them:LowerBound} are in fact optimizers of an intermediate inequality
(see Proposition \ref{prop:BastBBound} below) central to the proof
of Theorem \ref{them:CorrelationalBound}, so the results are directly
related. Furthermore, these states are natural generalizations of
the \textit{Yang pairing states} which Yang proved to precisely constitute
the optimizers of $\sup_{\Psi}\Vert\gamma_{2}^{\Psi}\Vert_{\mathrm{op}}$
in the \textit{finite}-dimensional setting.

\subsubsection*{A Conjecture for the Highly Correlated Regime}

We end this section by stating an explicit conjecture motivated by
the discrepancy between the two theorems: The strongest conjecture
one may consider is that for a fixed $\Phi\in\mathfrak{h}\wedge\mathfrak{h}$,
\begin{equation}
\sup_{\Psi\in\,\bigwedge^{N}\mathfrak{h},\,\left\Vert \Psi\right\Vert =1}\left\langle \Phi,\gamma_{2}^{\Psi}\Phi\right\rangle \leq\frac{N}{1+\frac{N-2}{2}\sum_{k=1}^{\infty}\lambda_{k}^{4}}.\label{eq:StrongConjecture}
\end{equation}
This can in general not hold, however, which can be seen as follows\footnote{I thank Robert Seiringer for pointing this out to me.}:
Let for $N$ even $\Psi_{N}$ denote the $N$-particle Yang pairing state based on $(u_{k})_{k=1}^{N}\cup(v_{k})_{k=1}^{N}$. It is well-known that $\gamma_{2}^{\Psi_{N}}$ has $N\mleft(1-\frac{N-2}{2N}\mright)=\frac{N}{2}+1$
as an eigenvalue, with the associated eigenvector being $\varphi_{N}=N^{-\frac{1}{2}}\sum_{k=1}^{N}u_{k}\wedge v_{k}$.
Thus
\begin{equation}
\left\langle \Phi,\gamma_{2}^{\Psi_{N}}\Phi\right\rangle \geq\mleft(\frac{N}{2}+1\mright)\left|\left\langle \varphi_{N},\Phi\right\rangle \right|^{2}=\mleft(\frac{N}{2}+1\mright)\mleft(\frac{1}{\sqrt{N}}\sum_{k=1}^{N}\lambda_{k}\mright)^{2}\geq\frac{1}{2}\mleft(\sum_{k=1}^{N}\lambda_{k}\mright)^{2}.
\end{equation}
Since there certainly are $\Phi\in\mathfrak{h}\wedge\mathfrak{h}$
for which $\sum_{k=1}^{\infty}\lambda_{k}=\infty$ we see that the
left-hand side of equation (\ref{eq:StrongConjecture}) can not admit
an $N$-independent bound for arbitrary $\Phi\in\mathfrak{h}\wedge\mathfrak{h}$.
The right-hand side is however bounded by $\sup_{N\in\mathbb{N}}\mleft(1+\frac{N-2}{2}\sum_{k=1}^{\infty}\lambda_{k}^{4}\mright)^{-1}N=\mleft(\frac{1}{2}\sum_{k=1}^{\infty}\lambda_{k}^{4}\mright)^{-1}$
uniformly in $N$, so the conjecture would yield a contradiction.\smallskip{}

It is central to this counterexample that we let $N\rightarrow\infty$,
so it does not limit the highly correlated regime $N\lambda_{\max}^{2}\ll1$.
Our conjecture is that in this case the asymptotics of Theorem \ref{them:LowerBound}
are indeed optimal:
\begin{conjecture}
Let $\Phi\in\mathfrak{h}\wedge\mathfrak{h}$ be normalized with canonical
form $\Phi=\sum_{k=1}^{\infty}\lambda_{k}u_{k}\wedge v_{k}$. Then
there exists a constant $C>0$, independent of $\Phi$, such that
for every $N\in2\mathbb{N}$ with $N\lambda_{\max}^{2}\leq1$
\[
\sup_{\Psi\in\,\bigwedge^{N}\mathfrak{h},\,\left\Vert \Psi\right\Vert =1}\left\langle \Phi,\gamma_{2}^{\Psi}\Phi\right\rangle \leq N\mleft(1-\frac{N-2}{2}\sum_{k=1}^{\infty}\lambda_{k}^{4}+C(N\lambda_{\max}^{2})^{2}\mright).
\]
\end{conjecture}

\section{Proof of Theorem \ref{them:CorrelationalBound}}

We first write $\Lambda=\left\langle \Phi,\gamma_{2}^{\Psi}\Phi\right\rangle $
in a particular manner: Employing the canonical form of $\Phi$ we
have
\begin{align}
\left\langle \Phi,\gamma_{2}^{\Psi}\Phi\right\rangle  & =\sum_{k,l=1}^{\infty}\lambda_{k}\lambda_{l}\left\langle u_{k}\wedge v_{k},\gamma_{2}^{\Psi}u_{l}\wedge v_{l}\right\rangle =2\sum_{k,l=1}^{\infty}\lambda_{k}\lambda_{l}\left\langle \Psi,c^{\ast}\mleft(u_{l}\mright)c^{\ast}\mleft(v_{l}\mright)c\mleft(v_{k}\mright)c\mleft(u_{k}\mright)\Psi\right\rangle \label{eq:PhiGammaPhiRewrite}\\
 & =2\left\langle \Psi,\mleft(\sum_{l=1}^{\infty}\lambda_{l}c^{\ast}\mleft(u_{l}\mright)c^{\ast}\mleft(v_{l}\mright)\mright)\mleft(\sum_{k=1}^{\infty}\lambda_{k}c\mleft(v_{k}\mright)c\mleft(u_{k}\mright)\mright)\Psi\right\rangle .\nonumber 
\end{align}
Motivated by the example of equation (\ref{eq:ExampleState}), let
us define
\begin{equation}
c_{k,\uparrow}^{\ast}=c^{\ast}\mleft(u_{k}\mright),\quad c_{k,\downarrow}^{\ast}=c^{\ast}\mleft(v_{k}\mright),
\end{equation}
for brevity. Then further defining
\begin{equation}
B=\sum_{k=1}^{\infty}\lambda_{k}c_{k,\downarrow}c_{k,\uparrow}
\end{equation}
we can write equation (\ref{eq:PhiGammaPhiRewrite}) simply as
\begin{equation}
\left\langle \Phi,\gamma_{2}^{\Psi}\Phi\right\rangle =2\left\langle \Psi,B^{\ast}B\Psi\right\rangle .
\end{equation}
The thing to note about the operator $B$ is that it behaves quasi-bosonically
with respect to its adjoint: We can compute
\begin{align}
\left[B,B^{\ast}\right] & =\sum_{k,l=1}^{\infty}\lambda_{k}\lambda_{l}\left[c_{k,\downarrow}c_{k,\uparrow},c_{l,\uparrow}^{\ast}c_{l,\downarrow}^{\ast}\right]=\sum_{k,l=1}^{\infty}\lambda_{k}\lambda_{l}\mleft(c_{k,\downarrow}\left\{ c_{k,\uparrow},c_{l,\uparrow}^{\ast}\right\} c_{l,\downarrow}^{\ast}-c_{l,\uparrow}^{\ast}\left\{ c_{k,\downarrow},c_{l,\downarrow}^{\ast}\right\} c_{k,\uparrow}\mright)\label{eq:BBastCommutator}\\
 & =\sum_{k=1}^{\infty}\lambda_{k}^{2}\mleft(c_{k,\downarrow}c_{k,\downarrow}^{\ast}-c_{k,\uparrow}^{\ast}c_{k,\uparrow}\mright)=1-\sum_{k=1}^{\infty}\lambda_{k}^{2}\mleft(c_{k,\uparrow}^{\ast}c_{k,\uparrow}+c_{k,\downarrow}^{\ast}c_{k,\downarrow}\mright)\nonumber 
\end{align}
and since the sum on the right-hand side obeys the bounds (with $\mathcal{N}$
denoting the usual number operator)
\begin{equation}
0\leq\sum_{k=1}^{\infty}\lambda_{k}^{2}\mleft(c_{k,\uparrow}^{\ast}c_{k,\uparrow}+c_{k,\downarrow}^{\ast}c_{k,\downarrow}\mright)\leq\lambda_{\max}^{2}\sum_{k=1}^{\infty}\mleft(c_{k,\uparrow}^{\ast}c_{k,\uparrow}+c_{k,\downarrow}^{\ast}c_{k,\downarrow}\mright)=\lambda_{\max}^{2}\mathcal{N}=N\lambda_{\max}^{2}
\end{equation}
on $\bigwedge^{N}\mathfrak{h}$ we see that $[B,B^{\ast}]\approx1$
in the highly correlated regime $N\lambda_{\max}^{2}\ll1$.

To motivate the next step, let us introduce the trial states which
we will use for Theorem \ref{them:LowerBound}: These are
\[
\Psi_{M}=(B^{\ast})^{M}\Omega,\quad M\in\mathbb{N},
\]
where $\Omega\in\bigwedge^{0}\mathfrak{h}\cong\mathbb{C}$ denotes
the vacuum vector. What is notable about these is that since
\begin{equation}
\left[c_{k,\uparrow},B^{\ast}\right]=\sum_{l=1}^{\infty}\lambda_{l}\left[c_{k,\uparrow},c_{l,\uparrow}^{\ast}c_{l,\downarrow}^{\ast}\right]=\sum_{l=1}^{\infty}\lambda_{l}\left\{ c_{k,\uparrow},c_{l,\uparrow}^{\ast}\right\} c_{l,\downarrow}^{\ast}=\lambda_{k}c_{k,\downarrow}^{\ast}
\end{equation}
and likewise $\left[c_{k,\downarrow},B^{\ast}\right]=-\lambda_{k}c_{k,\uparrow}^{\ast}$,
or in a unified notation
\begin{equation}
\left[c_{k,\sigma},B^{\ast}\right]=\pm_{\sigma}\lambda_{k}c_{k,\overline{\sigma}}^{\ast},\quad\mleft(\overline{\sigma},\pm_{\sigma}\mright)=\begin{cases}
\mleft(\downarrow,+\mright) & \sigma=\,\uparrow\\
\mleft(\uparrow,-\mright) & \sigma=\,\downarrow
\end{cases},
\end{equation}
the states $\Psi_{M}$ obey (as also $[[c_{k,\sigma},B^{\ast}],B^{\ast}]=0$)
\begin{equation}
c_{k,\sigma}\Psi_{M}=M\left[c_{k,\sigma},B^{\ast}\right](B^{\ast})^{M-1}\Omega=\pm_{\sigma}M\lambda_{k}c_{k,\overline{\sigma}}^{\ast}\Psi_{M-1}.\label{eq:cksigmaPsiM}
\end{equation}
Since $\Psi_{M}=B^{\ast}\Psi_{M-1}$ we can shift $M\rightarrow M+1$
and rearrange the resulting equation to see that
\begin{equation}
\mleft(\lambda_{k}c_{k,\sigma}^{\ast}\pm_{\sigma}\frac{1}{M+1}c_{k,\overline{\sigma}}B^{\ast}\mright)\Psi_{M}=0.\label{eq:PsiMEquation}
\end{equation}
Consideration of the operator in parenthesis will now lead us halfway
to Theorem \ref{them:CorrelationalBound}:
\begin{prop}
\label{prop:BastBBound}As an operator on $\bigwedge^{N}\mathfrak{h}$
it holds that
\[
B^{\ast}B\leq\frac{N}{2}-\frac{N-2}{4}\sum_{k=1}^{\infty}\lambda_{k}^{2}\mleft(c_{k,\uparrow}^{\ast}c_{k,\uparrow}+c_{k,\downarrow}^{\ast}c_{k,\downarrow}\mright).
\]
If the support of $(\lambda_{k})_{k=1}^{\infty}$ is infinite then
this inequality is optimal for every $N\in2\mathbb{N}$.
\end{prop}

\textbf{Proof:} By expansion we have for any $\alpha\in\mathbb{R}$
that
\begin{align}
\sum_{k=1}^{\infty}\sum_{\sigma\in\left\{ \uparrow,\downarrow\right\} }\left|\lambda_{k}c_{k,\sigma}^{\ast}\pm_{\sigma}\alpha c_{k,\overline{\sigma}}B^{\ast}\right|^{2} & =\sum_{k=1}^{\infty}\sum_{\sigma\in\left\{ \uparrow,\downarrow\right\} }\lambda_{k}^{2}c_{k,\sigma}c_{k,\sigma}^{\ast}+\sum_{k=1}^{\infty}\sum_{\sigma\in\left\{ \uparrow,\downarrow\right\} }\alpha^{2}Bc_{k,\overline{\sigma}}^{\ast}c_{k,\overline{\sigma}}B^{\ast}\\
 & +2\,\mathrm{Re}\sum_{k=1}^{\infty}\sum_{\sigma\in\left\{ \uparrow,\downarrow\right\} }\pm_{\sigma}\alpha\lambda_{k}c_{k,\sigma}c_{k,\overline{\sigma}}B^{\ast}.\nonumber 
\end{align}
We consider the right-hand side sum-by-sum. By the CAR the first can
be written as
\begin{equation}
\sum_{k=1}^{\infty}\sum_{\sigma\in\left\{ \uparrow,\downarrow\right\} }\lambda_{k}^{2}c_{k,\sigma}c_{k,\sigma}^{\ast}=2-\sum_{k=1}^{\infty}\lambda_{k}^{2}\mleft(c_{k,\uparrow}^{\ast}c_{k,\uparrow}+c_{k,\downarrow}^{\ast}c_{k,\downarrow}\mright)
\end{equation}
and since $\sum_{k=1}^{\infty}(c_{k,\uparrow}^{\ast}c_{k,\uparrow}+c_{k,\downarrow}^{\ast}c_{k,\downarrow})=\mathcal{N}$,
the second is
\begin{align}
\sum_{k=1}^{\infty}\sum_{\sigma\in\left\{ \uparrow,\downarrow\right\} }\alpha^{2}Bc_{k,\overline{\sigma}}^{\ast}c_{k,\overline{\sigma}}B^{\ast} & =\alpha^{2}B\mleft(\sum_{k=1}^{\infty}\mleft(c_{k,\uparrow}^{\ast}c_{k,\uparrow}+c_{k,\downarrow}^{\ast}c_{k,\downarrow}\mright)\mright)B^{\ast}\\
 & =\alpha^{2}B\mathcal{N}B^{\ast}=\mleft(N+2\mright)\alpha^{2}BB^{\ast}\nonumber 
\end{align}
when acting on $\bigwedge^{N}\mathfrak{h}$. Finally, upon expanding
the $\sigma$ sum the remaining part is seen to be
\begin{equation}
\sum_{k=1}^{\infty}\sum_{\sigma\in\left\{ \uparrow,\downarrow\right\} }\pm_{\sigma}\alpha\lambda_{k}c_{k,\sigma}c_{k,\overline{\sigma}}B^{\ast}=\alpha\sum_{k=1}^{\infty}\lambda_{k}\mleft(c_{k,\uparrow}c_{k,\downarrow}-c_{k,\downarrow}c_{k,\uparrow}\mright)B^{\ast}=-2\alpha BB^{\ast},
\end{equation}
so all in all there holds the identity
\begin{equation}
\sum_{k=1}^{\infty}\sum_{\sigma\in\left\{ \uparrow,\downarrow\right\} }\left|\lambda_{k}c_{k,\sigma}^{\ast}\pm_{\sigma}\alpha c_{k,\overline{\sigma}}B^{\ast}\right|^{2}=2-\alpha\mleft(4-\mleft(N+2\mright)\alpha\mright)BB^{\ast}-\sum_{k=1}^{\infty}\lambda_{k}^{2}\mleft(c_{k,\uparrow}^{\ast}c_{k,\uparrow}+c_{k,\downarrow}^{\ast}c_{k,\downarrow}\mright)
\end{equation}
on $\bigwedge^{N}\mathfrak{h}$. Optimizing $\alpha$ by setting $\alpha=\frac{2}{N+2}$
results in
\begin{equation}
\frac{4}{N+2}BB^{\ast}+\sum_{k=1}^{\infty}\sum_{\sigma\in\left\{ \uparrow,\downarrow\right\} }\left|\lambda_{k}c_{k,\sigma}^{\ast}\pm_{\sigma}\frac{2}{N+2}c_{k,\overline{\sigma}}B^{\ast}\right|^{2}=2-\sum_{k=1}^{\infty}\lambda_{k}^{2}\mleft(c_{k,\uparrow}^{\ast}c_{k,\uparrow}+c_{k,\downarrow}^{\ast}c_{k,\downarrow}\mright)
\end{equation}
and neglecting the sum on the left-hand side and multiplying by $\frac{N+2}{4}$
yields
\begin{equation}
BB^{\ast}\leq\frac{N+2}{2}-\frac{N+2}{4}\sum_{k=1}^{\infty}\lambda_{k}^{2}\mleft(c_{k,\uparrow}^{\ast}c_{k,\uparrow}+c_{k,\downarrow}^{\ast}c_{k,\downarrow}\mright).\label{eq:BBastBound}
\end{equation}
The statement of the proposition now follows by commuting $B$ and
$B^{\ast}$ (i.e. applying equation (\ref{eq:BBastCommutator})) and
rearranging the resulting expression.

The claim regarding optimality is a consequence of equation (\ref{eq:PsiMEquation}),
which shows that the term we neglected vanishes identically on $\Psi_{\frac{1}{2}N}$
(the assumption on the support enters to ensure that $\Psi_{\frac{1}{2}N}\neq0$).

$\hfill\square$

By this inequality we can conclude that
\begin{equation}
\left\langle \Phi,\gamma_{2}^{\Psi}\Phi\right\rangle =2\left\langle \Psi,B^{\ast}B\Psi\right\rangle \leq N-\frac{N-2}{2}\sum_{k=1}^{\infty}\lambda_{k}^{2}\mleft(\left\Vert c_{k,\uparrow}\Psi\right\Vert ^{2}+\left\Vert c_{k,\downarrow}\Psi\right\Vert ^{2}\mright)
\end{equation}
for \textit{any} normalized $\Psi\in\bigwedge^{N}\mathfrak{h}$. To
prove Theorem \ref{them:CorrelationalBound} it then only remains
for us to leverage the assumption that $\gamma_{2}^{\Psi}\Phi=\Lambda\Phi$
to control the quantities $\left\Vert c_{k,\sigma}\Psi\right\Vert ^{2}$
from below. We do this in the following:
\begin{prop}
Let $\Phi$ be an eigenvector of $\gamma_{2}^{\Psi}$ with eigenvalue
$\Lambda$. Then for every $k\in\mathbb{N}$
\[
\left\Vert c_{k,\uparrow}\Psi\right\Vert ^{2},\,\left\Vert c_{k,\downarrow}\Psi\right\Vert ^{2}\geq\frac{\Lambda}{2}\lambda_{k}^{2}.
\]
\end{prop}

\textbf{Proof:} Similarly to the approach of the previous proposition
we consider the square
\begin{align}
\left|c_{k,\sigma}\mp_{\sigma}\lambda_{k}c_{k,\overline{\sigma}}^{\ast}B\right|^{2} & =c_{k,\sigma}^{\ast}c_{k,\sigma}+\lambda_{k}^{2}B^{\ast}c_{k,\overline{\sigma}}c_{k,\overline{\sigma}}^{\ast}B\mp_{\sigma}2\lambda_{k}\,\mathrm{Re}\mleft(c_{k,\sigma}^{\ast}c_{k,\overline{\sigma}}^{\ast}B\mright)\\
 & =c_{k,\sigma}^{\ast}c_{k,\sigma}+\lambda_{k}^{2}B^{\ast}B-2\lambda_{k}\,\mathrm{Re}\mleft(c_{k,\uparrow}^{\ast}c_{k,\downarrow}^{\ast}B\mright)-\lambda_{k}^{2}B^{\ast}c_{k,\overline{\sigma}}^{\ast}c_{k,\overline{\sigma}}B.\nonumber 
\end{align}
As the left-hand side is manifestly non-negative and the last term
on the right-hand side is manifestly non-positive these can be neglected
and the resulting inequality rearranged for
\begin{equation}
c_{k,\sigma}^{\ast}c_{k,\sigma}\geq2\lambda_{k}\,\mathrm{Re}\mleft(c_{k,\uparrow}^{\ast}c_{k,\downarrow}^{\ast}B\mright)-\lambda_{k}^{2}B^{\ast}B.
\end{equation}
Now, note that (as in equation (\ref{eq:PhiGammaPhiRewrite}))
\begin{equation}
2\left\langle \Psi,c_{k,\uparrow}^{\ast}c_{k,\downarrow}^{\ast}B\Psi\right\rangle =\left\langle \Phi,\gamma_{2}^{\Psi}\mleft(u_{k}\wedge v_{k}\mright)\right\rangle ,
\end{equation}
so the assumption that $\gamma_{2}^{\Psi}\Phi=\Lambda\Phi$ (and self-adjointness
of $\gamma_{2}^{\Psi}$) implies
\begin{equation}
2\left\langle \Psi,c_{k,\uparrow}^{\ast}c_{k,\downarrow}^{\ast}B\Psi\right\rangle =\Lambda\left\langle \Phi,u_{k}\wedge v_{k}\right\rangle =\Lambda\lambda_{k}.
\end{equation}
As likewise $\left\Vert B\Psi\right\Vert ^{2}=\left\langle \Psi,B^{\ast}B\Psi\right\rangle =\frac{1}{2}\Lambda$
we thus find that
\begin{equation}
\left\Vert c_{k,\sigma}\Psi\right\Vert ^{2}\geq\lambda_{k}\,\mathrm{Re}\mleft(2\left\langle \Psi,c_{k,\uparrow}^{\ast}c_{k,\downarrow}^{\ast}B\Psi\right\rangle \mright)-\lambda_{k}^{2}\left\Vert B\Psi\right\Vert ^{2}=\lambda_{k}^{2}\Lambda-\frac{1}{2}\lambda_{k}^{2}\Lambda=\frac{\Lambda}{2}\lambda_{k}^{2}.
\end{equation}
$\hfill\square$

We can now conclude Theorem \ref{them:CorrelationalBound} as we have
obtained the estimate
\begin{equation}
\Lambda=\left\langle \Phi,\gamma_{2}^{\Psi}\Phi\right\rangle \leq N-\frac{N-2}{2}\sum_{k=1}^{\infty}\lambda_{k}^{2}\mleft(\left\Vert c_{k,\uparrow}\Psi\right\Vert ^{2}+\left\Vert c_{k,\downarrow}\Psi\right\Vert ^{2}\mright)\leq N-\frac{N-2}{2}\mleft(\sum_{k=1}^{\infty}\lambda_{k}^{4}\mright)\Lambda
\end{equation}
which yields the inequality of Theorem \ref{them:CorrelationalBound}
upon rearrangement.

\section{Proof of Theorem \ref{them:LowerBound}}

For Theorem \ref{them:LowerBound} we first note that since Proposition
\ref{prop:BastBBound} holds with equality on $\Psi_{M}$,
\begin{equation}
\left\langle \Psi_{M},B^{\ast}B\Psi_{M}\right\rangle =M\left\Vert \Psi_{M}\right\Vert ^{2}-\frac{M-1}{2}\sum_{k=1}^{\infty}\lambda_{k}^{2}\mleft(\left\Vert c_{k,\uparrow}\Psi_{M}\right\Vert ^{2}+\left\Vert c_{k,\downarrow}\Psi_{M}\right\Vert ^{2}\mright).
\end{equation}
Now, by the usual ``Pauli bound'' $\left\Vert c_{k,\sigma}\right\Vert _{\mathrm{op}}=1$
and equation (\ref{eq:cksigmaPsiM}) we can estimate
\begin{equation}
\left\Vert c_{k,\sigma}\Psi_{M}\right\Vert ^{2}=M^{2}\lambda_{k}^{2}\left\Vert c_{k,\overline{\sigma}}^{\ast}\Psi_{M-1}\right\Vert ^{2}\leq M^{2}\lambda_{k}^{2}\left\Vert \Psi_{M-1}\right\Vert ^{2}
\end{equation}
to conclude that the normalized states $\hat{\Psi}_{M}=\left\Vert \Psi_{M}\right\Vert ^{-1}\Psi_{M}$
obey
\begin{equation}
\left\langle \Phi,\gamma_{2}^{\hat{\Psi}_{M}}\Phi\right\rangle =2\left\langle \hat{\Psi}_{M},B^{\ast}B\hat{\Psi}_{M}\right\rangle \geq2M\mleft(1-\mleft(M-1\mright)\mleft(\sum_{k=1}^{\infty}\lambda_{k}^{4}\mright)\frac{M\left\Vert \Psi_{M-1}\right\Vert ^{2}}{\left\Vert \Psi_{M}\right\Vert ^{2}}\mright).
\end{equation}
It thus suffices to show that $\left\Vert \Psi_{M}\right\Vert ^{2}=M\left\Vert \Psi_{M-1}\right\Vert ^{2}\mleft(1+O(N\lambda_{\max}^{2})\mright)$,
which we obtain in the following:
\begin{prop}
For all $M\in\mathbb{N}$ it holds that
\[
\mleft(1-\mleft(M-1\mright)\lambda_{\max}^{2}\mright)M\left\Vert \Psi_{M-1}\right\Vert ^{2}\leq\left\Vert \Psi_{M}\right\Vert ^{2}\leq M\left\Vert \Psi_{M-1}\right\Vert ^{2}.
\]
\end{prop}

\textbf{Proof: }As also equation (\ref{eq:BBastBound}) holds with
equality for $\Psi_{M-1}$ we have
\begin{equation}
\left\Vert \Psi_{M}\right\Vert ^{2}=\left\langle \Psi_{M-1},BB^{\ast}\Psi_{M-1}\right\rangle =M\left\Vert \Psi_{M-1}\right\Vert ^{2}-\frac{M}{2}\sum_{k=1}^{\infty}\lambda_{k}^{2}\mleft(\left\Vert c_{k,\uparrow}\Psi_{M-1}\right\Vert ^{2}+\left\Vert c_{k,\downarrow}\Psi_{M-1}\right\Vert ^{2}\mright).
\end{equation}
The upper bound is immediate from this, and the lower bound follows
as
\begin{align}
\left\Vert \Psi_{M}\right\Vert ^{2} & \geq M\left\Vert \Psi_{M-1}\right\Vert ^{2}-\frac{M}{2}\lambda_{\max}^{2}\sum_{k=1}^{\infty}\mleft(\left\Vert c_{k,\uparrow}\Psi_{M-1}\right\Vert ^{2}+\left\Vert c_{k,\downarrow}\Psi_{M-1}\right\Vert ^{2}\mright)\\
 & =M\left\Vert \Psi_{M-1}\right\Vert ^{2}-\frac{M}{2}\lambda_{\max}^{2}\left\langle \Psi_{M-1},\mathcal{N}\Psi_{M-1}\right\rangle =M\left\Vert \Psi_{M-1}\right\Vert ^{2}\mleft(1-\mleft(M-1\mright)\lambda_{\max}^{2}\mright).\nonumber 
\end{align}
$\hfill\square$

(We remark that with further analysis one can show that in fact
\begin{equation}
\left\Vert \Psi_{M}\right\Vert ^{2}=M\left\Vert \Psi_{M-1}\right\Vert ^{2}\mleft(1-\mleft(M-1\mright)\sum_{k=1}^{\infty}\lambda_{k}^{4}+O(N^{2}\lambda_{\max}^{4})\mright)
\end{equation}
i.e. the first-order correction in the highly correlated regime is
again determined solely by $\sum_{k=1}^{\infty}\lambda_{k}^{4}$.)

If $M\lambda_{\max}^{2}\leq\frac{1}{2}$ (say) we can thus estimate
\begin{equation}
\frac{M\left\Vert \Psi_{M-1}\right\Vert ^{2}}{\left\Vert \Psi_{M}\right\Vert ^{2}}\leq\frac{1}{1-\mleft(M-1\mright)\lambda_{\max}^{2}}=1+\frac{\mleft(M-1\mright)\lambda_{\max}^{2}}{1-\mleft(M-1\mright)\lambda_{\max}^{2}}\leq1+2M\lambda_{\max}^{2}
\end{equation}
and thereby conclude that the normalized states $\hat{\Psi}_{M}$
obey
\begin{equation}
\left\langle \Phi,\gamma_{2}^{\hat{\Psi}_{M}}\Phi\right\rangle \geq2M\mleft(1-\mleft(M-1\mright)\sum_{k=1}^{\infty}\lambda_{k}^{4}-2M^{2}\lambda_{\max}^{4}\mright),\quad M\lambda_{\max}^{2}\leq\frac{1}{2}.
\end{equation}
For any even $N$ with $N\lambda_{\max}^{2}\leq1$ the choice $\Psi=\hat{\Psi}_{\frac{1}{2}N}$
thus verifies Theorem \ref{them:LowerBound}.

\end{document}